\documentclass[12pt]{article} 
\pdfoutput=1
\usepackage{amssymb,graphicx}
\usepackage{epstopdf}
\usepackage{amsmath,amsfonts}
\usepackage{epsfig} 
\usepackage{graphicx,graphics}
\usepackage{subcaption}
\usepackage{tikz}
\usepackage[dvipsnames]{xcolor}
\usepackage{hyperref}

\begin{document} 

\title{\bf Natural saturation of the sterile neutrino dark matter resonant production by a high lepton flavor asymmetry in primordial plasma}
\author{D.~Gorbunov$^{a, b}$, D.~Kalashnikov$^{a, b}$\\
\small{\em $^a$Institute for Nuclear Research of the Russian Academy of Sciences, 117312 Moscow, Russia}\\
\small{\em $^b$Moscow Institute of Physics and Technology, 141700 Dolgoprudny, Russia}\\
}
 
 \date{}

{\let\newpage\relax\maketitle}

\begin{abstract}

Sterile neutrinos remain a well-motivated dark-matter candidate whose cosmological abundance can be enhanced by resonant active--sterile conversion in the presence of a lepton asymmetry in primordial plasma. In the standard picture, a larger initial asymmetry increases the matter potential and can therefore promote resonant production. However, the same increase also shifts the resonance to later stages of the cosmological evolution when the plasma temperature is lower. We show that this delayed production epoch can overlap with the onset of active neutrino oscillations, which redistribute the flavor asymmetries and may substantially reduce, or in some cases nearly erase, the lepton asymmetry needed for a successful resonant sterile-neutrino production. This interplay provides a natural saturation mechanism for the final sterile-neutrino abundance: beyond a certain range of initial asymmetries, increasing the primordial asymmetry no longer leads to a proportional increase in the produced dark-matter density. We identify this effect as an additional constraint on resonant sterile-neutrino production at large lepton asymmetry and discuss its dependence on the flavor structure of the initial asymmetry and on the background cosmological evolution. Our numerical results reveal two order of magnitude range in sterile-active neutrino mixing parameter $\sin^2(2\theta)$ presently consistent with the resonant mechanism of the dark matter sterile neutrino production. It must be investigated by the next generation X-ray telescopes to fully explore this mechanism and corresponding minimal  models suggesting sterile neutrinos as viable dark matter. We also consider pion condensation at large flavor lepton asymmetries, which may lead to a first-order QCD phase transition and associated gravitational-wave production in the early Universe.
\end{abstract}

\newpage

\section{Introduction}
\label{sec:intro}

Sterile neutrinos provide a well-motivated extension of the Standard Model and remain viable candidates for cosmological dark matter, for reviews see e.g. \cite{Drewes:2016upu,Boyarsky:2018tvu}. If their masses are in the keV range and their mixing with active neutrinos is sufficiently small, they can be long-lived on cosmological timescales while still being produced in the early Universe through active--sterile neutrino oscillations. The simplest production mechanism is (non-resonant) thermal production, also known as the Dodelson--Widrow mechanism, in which sterile states are gradually populated by oscillations and collisions in the primordial plasma \cite{Dodelson:1993je}. Extensive searches for the peak signature, predicted in the galaxy $X$-ray spectra\,\cite{Abazajian:2001vt}, have excluded the parameter region in which this mechanism alone produces the entire observed dark-matter abundance, for most recent experimental limits see 
Refs.\,\cite{Foster:2021ngm,Roach:2022lgo,Zakharov:2023mnp,Krivonos:2024yvm}. A qualitatively different situation arises in the presence of a nonzero lepton asymmetry in the primordial plasma. In this case, the asymmetry modifies the finite-density matter potential experienced by active neutrinos and can induce the resonant   active--sterile conversion \cite{Shi:1998km}. This resonant production, known in literature as the Shi--Fuller mechanism, is therefore especially important because it can strongly affect both the final sterile-neutrino abundance and the momentum distribution of the produced dark-matter population\,\cite{Abazajian:2001nj,Laine:2008pg}.

The size and flavor structure of the primordial lepton asymmetry are constrained by cosmological observables, most notably by the Big Bang Nucleosynthesis (BBN), which severely limits the asymmetry between electron neutrino and antineutrino   \cite{Escudero:2022okz,Froustey:2024mgf}. These constraints are, however, not determined solely by the initial asymmetries assigned to the individual lepton flavors. Before BBN, active neutrino oscillations can redistribute the asymmetry among flavors and change the relation between the initial conditions and the asymmetries relevant for light-element production \cite{Dolgov:2002ab,Barenboim:2016shh,Gorbunov:2025nqs,Domcke:2025lzg}. In some regions of parameter space, the oscillation dynamics may substantially reduce, or even nearly erase, initially large flavor asymmetries before BBN. A consistent treatment of lepton-asymmetric cosmologies must therefore account for the flavor evolution of the active-neutrino sector rather than impose bounds directly on the initial asymmetry parameters.

The maximal parameter space of sterile-neutrino dark matter in the presence of lepton-flavor asymmetries was recently investigated in Ref.~\cite{Akita:2025txo}. That work studied how large an abundance can be obtained through resonant production when the asymmetry is distributed non-trivially among the active flavors. Its treatment is nevertheless restricted to asymmetries, characterized by the ratios of lepton flavor number densities to entropy density, of roughly $L<0.1$. This limitation arises because two physical effects become important at larger asymmetries: the possible onset of pion condensation and the dynamical depletion of the lepton asymmetry by active neutrino oscillations. Without these ingredients, extrapolating the calculation of sterile-neutrino abundance to larger initial flavor asymmetries is at least incomplete.

The present work combines and extends these two directions. We continue our previous study of resonant sterile-neutrino dark-matter production with nontrivial lepton-flavor asymmetries \cite{Gorbunov:2025nqs}, in which we examined how different distributions of the total asymmetry among flavors affect sterile-neutrino production and cosmological constraints. Here we go beyond that treatment by including pion condensation and active-neutrino oscillations in the evolution of the lepton asymmetries used in the sterile-neutrino production calculation. This extension is necessary because the matter potential that controls the resonance is itself determined by the evolving flavor asymmetries and by the charged background of the primordial plasma.

For numerical implementations, we apply and adapt the tools developed to perform the sterile-neutrino abundance calculation \cite{Akita:2025txo} and the tools to perform the evaluation of active-neutrino oscillations \cite{Domcke:2025lzg} in the primordial plasma with lepton-flavor asymmetries. We combine the sterile-neutrino production code with the active-neutrino oscillation code, modify both for the present problem, and include a description of the cosmological background during the QCD epoch at large lepton asymmetry. In particular, we incorporate pion condensation following Ref.~\cite{Ferreira:2025zeu}, which modifies the charge-neutrality condition and the chemical potentials entering the matter potential.

The main physical effect identified in this work is the following. Increasing the initial lepton asymmetry shifts resonant sterile-neutrino production to later cosmological times, or equivalently to lower temperatures. At sufficiently late times, active neutrino oscillations become efficient and can deplete the lepton asymmetry before the expected or during the epoch of sterile-neutrino production. Consequently, the final sterile-neutrino abundance does not necessarily grow indefinitely with the initial asymmetry. Instead, in relevant regions of parameter space, the abundance exhibits a saturation behavior: increasing the primordial asymmetry further no longer leads to a corresponding increase in the final sterile-neutrino dark-matter density.

This paper is organized as follows. In Secs.~\ref{Sec:cond}, \ref{Sec:cond2} we give a detailed description of the cosmological background at large lepton asymmetry, including the pion-condensed phase. In Sec.~\ref{Sec:active} we discuss the implementation of active neutrino oscillations and their impact on the evolution of the background asymmetries. In Sec.\,\ref{Sec:prod} we discuss the analytical approximations adopted to calculate numerically the spectra and the amount of resonantly produced sterile neutrinos. 
In Secs. \ref{Sec:asym}, \ref{Sec:abund} we then performed the calculation of the evolution of lepton asymmetries and sterile-neutrino production and analyzed the resulting abundance for the representative set of initial flavor-asymmetry configurations, highlighting the saturation mechanism induced by active-oscillation depletion. Finally, in Sec.\,\ref{Sec:minimal-mixing} we extrapolate out results for these configurations to outline the a factor-of-two estimate of the minimal mixing angle to produce the sterile neutrinos at the amount sufficient to fully explain the dark matter component of the Universe. In Sec.\,\ref{sec:sterile_spectra} we presented examples of typical spectra of the relic sterile neutrinos we obtained, which exhibit various features. They illustrate the importance of further increasing the accuracy of numerical calculations needed to get the robust spectra sterile neutrino dark matter to be used in simulations of cosmic large scale structure formation.  

\section{Cosmological background evolution}
The calculation of the sterile-neutrino abundance requires the thermodynamic background of the early Universe to be closely followed throughout the temperature interval in which resonant production occurs. Increasing the initial lepton asymmetry shifts the resonance toward lower temperatures; for the parameter range considered here, the interesting evolution takes place predominantly at temperatures of $T \simeq 5$--$150\,\mathrm{MeV}$. Thus we determine the energy density, entropy density, expansion rate, and matter potential self-consistently across this interval. In the adopted description, the hadronic contribution is represented by nucleons and light mesons, while the lepton asymmetries are encoded in flavor-dependent chemical potentials constrained by charge neutrality with respect to the electric charge.

The repeated evaluation of this background is one of the most computationally demanding components of the numerical analysis, particularly when the active neutrinos rapidly oscillate in the plasma. It is therefore useful to identify controlled simplifications that reduce the numerical cost without compromising the precision required for the convincing calculation of sterile-neutrino abundance. In this section, we formulate the system of equations describing the evolution of chemical potentials, justify the disregard of the baryon chemical potential, and then discuss the additional effects of pion condensation and active flavor conversion.

\subsection{Description of the chemical potentials and
baryons}
\label{Sec:cond}

We evaluate the complete thermodynamic system following the approach and adapting the numerical code of Ref.~\cite{Akita:2025txo}. The system contains five independent chemical potentials: $\mu_{\nu_e}$, $\mu_{\nu_\mu}$, $\mu_{\nu_\tau}$ for active neutrino species, and $\mu_Q$ for electric charge, and $\mu_B$ for baryon charge. The remaining variable for lepton charges and individual hadron (proton, neutron and charged pion) chemical potentials are related to them by
\begin{align}
  \mu_{l_\alpha} &= \mu_{\nu_\alpha}-\mu_Q, \\
  \label{proton}
  \mu_p &= \mu_B+\mu_Q, 
  \\
  \label{neutron}
  \mu_n &= \mu_B, \\
  \label{pion}
  \mu_\pi &= -\mu_Q.
\end{align}
These relations are fixed in the plasma by the three flavor conditions, electric charge neutrality, and the observed baryon asymmetry.

Remarkably, at the temperatures relevant for the present calculation, the proton and neutron densities are strongly suppressed by the nucleon masses. Within our treatment of the hadronic sector, see eqs.\eqref{proton},\eqref{neutron}, $\mu_B$ affects the background only through these densities. We therefore set $\mu_B=0$ and omit the baryon-number equation, reducing the problem to four chemical potentials, $\mu_{\nu_e}$, $\mu_{\nu_\mu}$, $\mu_{\nu_\tau}$, and $\mu_Q$. The reduced system for the corresponding number densities is
\begin{align} \label{eq:BG_chem}
  n_{L_\alpha} &= L_\alpha s, \qquad \alpha=e,\mu,\tau, \\
  n_Q &= 0,
\end{align}
where
\begin{equation}
  L_\alpha\equiv\frac{\Delta n_{\nu_\alpha}+\Delta n_{l_\alpha}}{s}\,,
\end{equation}
with differences of particle and antiparticle number densities $\Delta n_f\equiv n_f-n_{\bar f}$. 
These equations determine the chemical potentials and consequently the background energy density and entropy density $s$ at each temperature $T$ assuming chemical (kinetic) equilibrium in the plasms.

We have checked the accuracy of this approximation by comparing the referenced results obtained with the full and with the the reduced system (i.e. with zero baryon charge). Over the temperatures and asymmetries considered in this work, omitting $\mu_B$ changes the relevant background quantities by less than $1\%$. The baryon sector can therefore be neglected at the precision required for the sterile-neutrino abundance calculation, while substantially simplifying the numerical solution. The baryon chemical potential also enters the description of evolution of the pion-condensed phase; as we discuss below in Sec.~\ref{Sec:cond}, its effect can be neglected there as well.

\subsection{Pion condensation} 
\label{Sec:cond2}

At sufficiently large lepton asymmetries, the electric charge chemical potential can approach the charged-pion mass. The sign of $\mu_Q$ selects which charged-pion state is enhanced; equivalently, the chemical potential of the favored pion species approaches $\mu_\pi=m_\pi$ when $|\mu_Q|=m_\pi$, see eq.\,\eqref{pion}. A Bose--Einstein chemical potential cannot exceed the particle mass. As this limiting value is approached, the occupation of the low-momentum modes becomes singular, and any additional charge must be accommodated by a zero-momentum condensate. We describe this phase using the quark--meson model of Ref.~\cite{Ferreira:2025zeu}.

The charged-pion number density is given by integrating the pion phase space density over 3-momentum $p$
\begin{equation} 
    n_\pi = \frac{1}{(2\pi)^3}\int d^3p\, f_\pi(p,\mu_\pi,T)
    =\frac{1}{2\pi^2}\int_0^\infty dp\,p^2 f_\pi(p,\mu_\pi,T).
\end{equation}
In the condensed phase we set $\mu_\pi=m_\pi$ and separate the thermal distribution from the zero-momentum contribution \cite{Dolgov:2008pe}:
\begin{equation}
\label{pion-spectrum}
    f_\pi(p,m_\pi,T)=f_B(p,m_\pi,T)+C(\mu_Q,T)\delta^{(3)}(\vec{p}),
\end{equation}
where
\begin{equation}
\label{BE}
    f_B(p,\mu_\pi,T)\equiv
    \frac{1}{\exp\!\left[\left(\sqrt{p^2+m_\pi^2}-\mu_\pi\right)/T\right]-1}.
\end{equation}
For $p/T\ll1$ and $p/m_\pi\ll1$, this distribution at $\mu_\pi=m_\pi$ reduces to
\begin{equation}
\label{low}
    f_B(p,m_\pi,T)\simeq\frac{2m_\pi T}{p^2}.
\end{equation}
In the numerical calculations we approximate the pion spectra by splitting the 3-momentum space at $p_s=0.01T$. For $p>p_s$ we utilize the complete Bose--Einstein distribution\,\eqref{BE}, whereas for $0<p<p_s$ we use the low-momentum expression\,\eqref{low}. The coefficient of the zero-momentum term in eq.\,\eqref{pion-spectrum} is related to the pion-condensate amplitude by
\begin{equation}\label{eq:Cond}
    C(\mu_Q,T)=(2\pi)^3\frac{1}{2}\Pi_0^2(\mu_Q,T)m_\pi
    \equiv (2\pi)^3m_\pi\pi_0^2(\mu_Q,T),
\end{equation}
where $\Pi_0$ is the amplitude of the condensed field and $\pi_0=\Pi_0/\sqrt{2}$ is its mean value. The complete charged-pion density is therefore
\begin{equation}
    n_\pi=\frac{m_\pi T}{\pi^2}p_s
    +\frac{1}{2\pi^2}\int_{p_s}^\infty dp\,p^2 f_B(p,m_\pi,T)
    +m_\pi\pi_0^2(\mu_Q,T),
\end{equation}
with $C/(2\pi)^3=m_\pi\pi_0^2$. For the temperatures considered here, the choice $p_s=0.01T$ also ensures $p_s/m_\pi\ll1$, so that the low-momentum approximation is always valid throughout the integration over momenta from zero to $p_s$.

To determine $\pi_0$, we use the two-flavor quark--meson model and minimize its thermodynamic potential with respect to the chiral and pion condensates, $(\sigma_0,\pi_0)$, while simultaneously imposing the cosmological charge-neutrality condition $n_Q=0$. The thermodynamic potential $\Omega$ has two parts: fermionic potential  
\begin{equation}
\begin{split}
    \Omega_F = - 6T\int\frac{d^3p}{(2\pi)^3} \times & \left[ \ln \left( 1 + e^{(-E_\Delta^-+\mu_q)/T} \right) + \ln \left( 1 + e^{(-E_\Delta^--\mu_q)/T} \right) \right. \\ & \left. + \ln \left( 1 + e^{(-E_\Delta^++\mu_q)/T} \right) + \ln \left( 1 + e^{(-E_\Delta^+-\mu_q)/T} \right) \right],
\end{split}
\end{equation}
with
\begin{equation}
    E_\Delta^\pm = \sqrt{(E_q\pm\mu_I)^2 + \Delta^2},
\end{equation}
and mesonic potential
\begin{equation}
    \Omega_M = \frac{\lambda}{4}(\sigma_0^2 + \pi_0^2 - v^2)^2 - h\sigma_0 - \frac{1}{2}\mu_Q^2 \pi_0^2,
\end{equation}
where $E_q = \sqrt{p^2+m_q^2}$, $m_q=g\sigma_0$ and $\Delta=g\pi_0$, $\mu_I=\mu_Q/2$, $\mu_q=\mu_B/3+\mu_Q/6$. Following Ref.~\cite{Ferreira:2025zeu}, we fix the parameters $g$, $\lambda$, $h$, and $v$ in the vacuum, at $\mu=0$, $\sigma_0=f_\pi$, and $\pi_0=0$, so as to reproduce the pion decay constant $f_\pi=92\,\mathrm{MeV}$ and meson masses $m_\pi=138\,\mathrm{MeV}$, $m_\sigma=700\,\mathrm{MeV}$:

\begin{equation}
\label{eq:QM_param}
\begin{cases}
    \lambda(f_\pi^2 - v^2)f_\pi - h = 0 \\
    m_\sigma^2 = \lambda(3f_\pi^2-v^2) \\
    m_\pi^2 = \lambda(f_\pi^2-v^2) \\
    m_q = g f_\pi
\end{cases} 
\Rightarrow
\begin{cases}
    \lambda = \frac{m_\sigma^2-m_\pi^2}{2f_\pi^2} = 27.8\\
    v = \sqrt{f_\pi^2 - \frac{m_\pi^2}{\lambda}} = 88.2 \; \text{MeV}\\
    h = \lambda f_\pi (f_\pi^2-v^2) = (120.6 \; \text{MeV})^3\\
    g = \frac{m_q}{f_\pi} \sim 3.26
\end{cases}
\end{equation}

For the large asymmetries of primary interest here, the fermionic contribution to $\Omega$ becomes subdominant. Omitting this term simplifies the condensate calculation to analytical solution, where $\mu_B$ does not enter the equations. And the condensate contribution to the charge density becomes
\begin{equation}
    \Delta n_Q^{(\text{cond})}
    =-\frac{\partial\Omega}{\partial\mu_Q}
    =\mu_Q\pi_0^2.
\end{equation}

\begin{figure}[htbp!]
  \centering
    \includegraphics[width=0.8\textwidth]{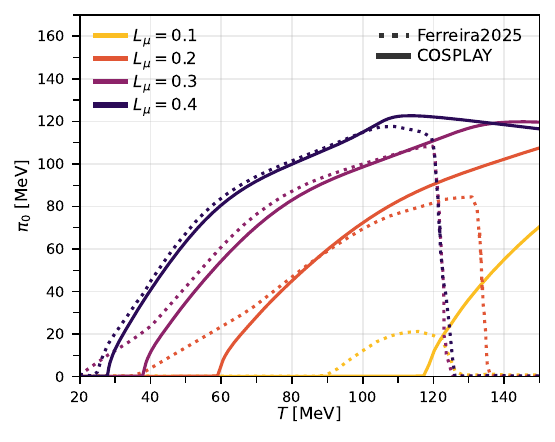}
  \caption{Condensate amplitudes along cosmological trajectories as functions of temperature. Dotted lines show the full quark--meson calculation of Ref.~\cite{Ferreira:2025zeu}, while solid lines show the simplified treatment used in this work.}
  \label{fig:Condensate}
\end{figure}

Figure~\ref{fig:Condensate} 
shows the comparison of the results obtained applying   our simplified prescription with the condensate amplitudes reported in Ref.~\cite{Ferreira:2025zeu}; the corresponding amplitudes are already close to each other for $L=0.4$. The high-temperature difference reflects our usage of a pion-based description throughout the interval: for the trajectories shown, $|\mu_Q|>m_\pi$ already at $T=150\,\mathrm{MeV}$, so the condensate is present at the initial temperature. We do not attempt to interpret this high-temperature effect further, because for the larger asymmetries relevant to our main analysis the sterile-neutrino resonances occur below $100\,\mathrm{MeV}$. To estimate errors introduced by the simplified pion condensate evaluation we conducted additional sensitivity tests. They show that, at lower asymmetries, $L=0.2-0.3$ substituting the condensate amplitudes from Ref.~\cite{Ferreira:2025zeu} can change $\Omega_{\nu_s}$ by roughly $5\%$ at the higher tested sterile neutrino masses. While for the ligher neutrino this discrepancy falls below 1\% because the lighter neutrinos enter resonance at later times, where the plasma temperature in the expanding Universe drops and the condensate is already disappeared. For the asymmetry $L_\mu=1$ and sterile neutrino mass $m_s=100\,\mathrm{keV}$, the simplified-condensate and zero-condensate calculations differ by approximately $20\%$ in $\Omega_{\nu_s}$. 

The physical motivation for the only meson approximation follows from the minimizing the thermodynamic potential. With $E_q=\sqrt{p^2+g^2\sigma_0^2}$, differentiation with respect to pion field at fixed temperature and chemical potentials gives
\begin{equation}\label{eq:pion_gap_equation}
\begin{split}
    \frac{\partial\Omega}{\partial\pi_0}
    ={}&\pi_0\biggl[\lambda(\sigma_0^2+\pi_0^2-v^2)-\mu_Q^2\\
    &+6g^2\int\frac{d^3p}{(2\pi)^3}\sum_{a=\pm}
    \frac{f_F(E_\Delta^a-\mu_q)+f_F(E_\Delta^a+\mu_q)}{E_\Delta^a}\biggr],
\end{split}
\end{equation}
where $f_F(x)=(e^{x/T}+1)^{-1}$. The integral is the fermionic correction to the simple only meson model. In the limit $E_\Delta^a\mp\mu_q \gg T$, the distribution functions vanish, $f_F \approx 0$. 
A sufficient condition for exponentially small occupations reads
\begin{equation}\label{eq:pion_thermal_suppression}
    \frac{E_\Delta^a-|\mu_q|}{T}\gg1.
\end{equation}
The minimum quasiparticle energy is
\begin{equation}\label{eq:pion_quasiparticle_gap}
\begin{split}
    E_{\rm gap} = \min_{p,a}E_\Delta^a
    =\sqrt{\Delta^2+\left[\max(m_q-|\mu_I|,0)\right]^2}
    \geq|\Delta|.
\end{split}
\end{equation}
For $g\simeq3.26$, the pion condensate field of $40$--$100$ MeV gives $|\Delta|\simeq132$--$326$ MeV. Such gaps can suppress the thermal distributions at $T\sim 10$ MeV. 

The baryon chemical potential enters the system through the quark chemical potential $\mu_q=\mu_B/3+\mu_Q/6$. In a baryon-inclusive diagnostic calculation with fixed initial asymmetries $L_\mu=1$, $L_e=0$, and $L_\tau=-1$, using the simplified condensate, $\mu_B$ tends to be $\simeq -\mu_Q/2$. This keeps $\mu_q$ close to zero at higher temperatures. For example, at $T\simeq 50$\,MeV, one gets $\mu_q\simeq-0.14$\,MeV and $E_{\rm gap}\simeq400$\,MeV. At temperature below approximately $T=40$\,MeV, $\mu_B$ and consequently $\mu_q$ rises rapidly. At $T=20$\,MeV, one finds $\mu_q\simeq144$\,MeV and $E_{\rm gap}\simeq248.9$\,MeV. The suppression ratio is then approximately $5.2\times10^{-3}$. Suppression may weaken if $|\mu_q|$ approaches the gap. For $|\mu_q|>E_{\rm gap}$ the exponential argument fails.

Repeatedly solving the full condensate and chemical-equilibrium system during active-flavor evolution would add substantial computational cost. We retain the simplified mesonic prescription for the present analysis, supported by the reference comparison and the benchmark sensitivity tests, while allowing model dependence at the several-percent to tens-of-percent level in the tested cases.

\subsection{Active-neutrino oscillations} \label{Sec:active}
As the Universe expands, at temperatures of order $20\,\mathrm{MeV}$ and below the active-neutrino oscillations begin to redistribute the asymmetries among the three flavors. Both the precise moment when the oscillations start and the efficiency of this conversion depend on the initial flavor configuration and on the matter potential. Although active oscillations conserve the total lepton asymmetry, they can strongly reduce the asymmetry of the particular flavor, which enters the corresponding active--sterile matter potential. Their effect must therefore be accounted simultaneously with the thermodynamic background and the lepton-asymmetry evolution.

We use a modified version of COFLASY \cite{Domcke:2025lzg} to calculate the active-neutrino flavor evolution by solving the quantum kinetic equations for the neutrino density matrices. The oscillation calculation itself is unchanged: we do not modify the Hamiltonian, collision terms, momentum treatment, integration procedure, or any other routine that evaluates the neutrino evolution. Our modifications are restricted to the input, output, and driver sections of the code. These changes allow COFLASY to exchange the flavor asymmetries and the required thermodynamic quantities with the adapted Python code of \cite{Ferreira:2025zeu}, exploited to evaluate the cosmological background.

\begin{figure}[ht!]
    \centering
    \begin{tikzpicture}[>=stealth, line width=0.9pt]
        \draw[<-] (0,0) node[left] {$T$} -- (10,0) ;
        \draw (2,0.18) -- (2,-0.18);
        \draw (7,0.18) -- (7,-0.18);

        \node[below] at (2,-0.18) {$T_i$};
        \node[below] at (7,-0.18) {$T_{i+1}$};

        \draw[->] (2,2.2) -- (2,0.28);
        \draw[->] (7,2.2) -- (7,0.28);
        \node[left] at (2,1.35) {$L_\alpha^i$};
        \node[right] at (7,1.35) {$L_\alpha^{i+1}=L_\alpha^i + \delta L_\alpha^i$};

        \draw[->, bend left=40] (2.25,0.2) to node[midway, above=13pt, align=center] {active-flavor evolution} (6.75,0.2);
        \node[above] at (4.5,0.95) {COFLASY};
    \end{tikzpicture}
    \caption{Schematic illustration of one step in the joint evolution of the coupled system. The background solver provides the thermodynamic state at the temperature $T_i$, COFLASY evolves the active-neutrino sector to the next temperature $T_{i+1}$, and the resulting flavor-asymmetry
    update is returned to the background calculation.}
    \label{fig:osc_scheme}
\end{figure}
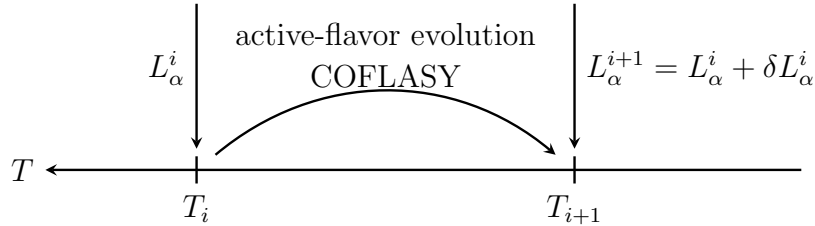

The joint evolution of the coupled system is performed on a decreasing temperature grid $T_0>T_1>\cdots>T_N$. At a grid point $T_i$, the current flavor asymmetries $L_\alpha^i$ are transferred to the background solver. Solving the background equations \eqref{eq:BG_chem} determines the chemical potentials, entropy density, and other thermodynamic quantities required over the next temperature interval. These data, together with the neutrino state at $T_i$, are passed to COFLASY, which calculate the evolution of the active-neutrino density matrices from $T_i$ to $T_{i+1}$. The flavor asymmetries extracted at the end of this evolution define the shifts in the asymmetries $\delta L_\alpha^i$ and hence the values of the flavor asymmetries at the next grid point, 
\begin{equation}
    L_\alpha^{i+1}=L_\alpha^i+\delta L_\alpha^i.
\end{equation}
The updated asymmetries are then returned to the background solver at $T_{i+1}$, and the procedure is repeated. In this way, the flavor conversion changes the chemical potentials and matter potentials used at the subsequent steps, while the evolving background consistently determines the (averaged) conditions under which the next interval of active oscillations takes place. The information flow during one step is summarized in Fig.~\ref{fig:osc_scheme}.

\begin{figure}[htb!]
    \centerline{
        \includegraphics[width=0.5\textwidth]{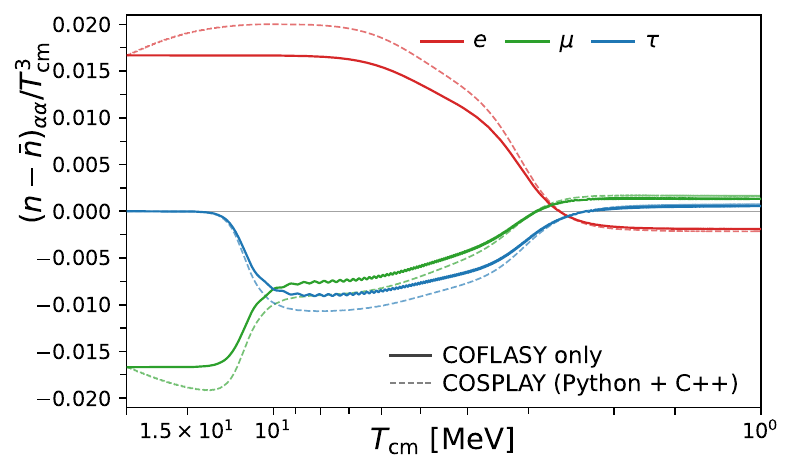}
        \includegraphics[width=0.5\textwidth]{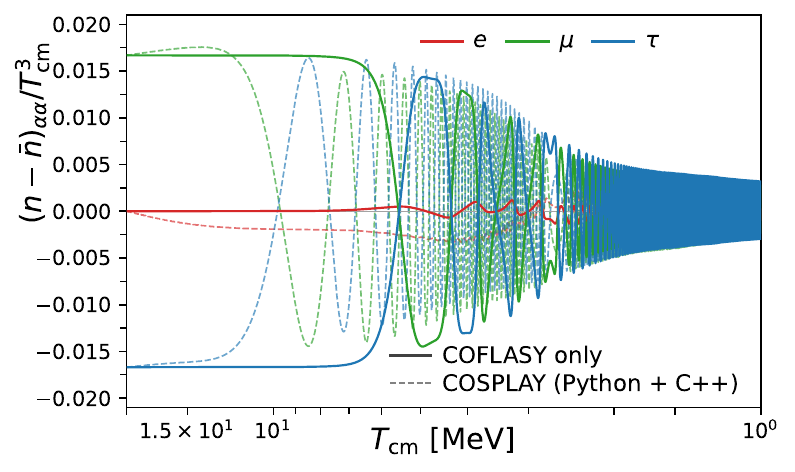}}
    \caption{Comparison of the flavor-asymmetry evolution obtained with the standalone C++ version of COFLASY (solid curves) and with the complete temperature-stepping framework used in this work (dashed curves), for two representative initial flavor configurations. The modified C++ code closely reproduces the original standalone result for identical neutrino initial conditions.}
    \label{fig:coflasy_validation}
\end{figure}

We first verified that our modified C++ implementation reproduces the output of the original standalone COFLASY code when both are initialized with identical neutrino distributions. Figure~\ref{fig:coflasy_validation} 
instead compares the standalone calculation with the complete temperature-stepping scheme used in this work. The differences between the two evolutions have a physical origin. At temperatures around $20\,\mathrm{MeV}$ and for large flavor asymmetries, charged leptons still make a non-negligible contribution to the total lepton flavor asymmetries, with the muon contribution being particularly important. As the Universe cools and the muons disappear from the plasma, this asymmetry is transferred to the neutrino sector. The neutrino asymmetries that enter the subsequent active-oscillation epoch are therefore slightly shifted relative to those imposed directly in the standalone calculation. The resulting difference between the curves reflects the self-consistent background evolution in the stepped scheme rather than a modification of the C++ oscillation dynamics.

The low-temperature Python--COFLASY interface supports two interaction regimes which we call the continuation and the fresh-step initialization. In the continuation mode, the Python code supplies the C++ solver with the complete replica of the dynamical state from the last successful interval: the 20-component vector $(y_0,\ldots,y_{19})$, the original temperature scale $T_{\mathrm{scale}}$, and the final dimensionless evolution time $t_{\mathrm{final}}$, which becomes the starting time of the next calculation interval. The state vector contains 18 entries of the neutrino and antineutrino density matrices in the Gell-Mann decomposition, together with the neutrino and photon background variables. Python additionally supplies the three flavor-density asymmetries newly reconstructed by the background solver. Before resuming the ordinary differential equation integration, COFLASY adjusts the diagonal elements of the neutrino and antineutrino density matrices to reproduce these asymmetries, while retaining the remaining information in the saved state, including its off-diagonal coherence structure and temperature variables. In contrast, the fresh-step initialization discards the previous 20-component state: the Python code passes only the three flavor-density asymmetries and the initial and final temperatures of the interval. COFLASY converts these asymmetries into degeneracy parameters, constructs a new equilibrium-like initial density-matrix state, and evolves it across the specified temperature interval. In the comparisons performed so far, the solutions obtained with the fresh-step initialization appear to numerically follow the expected adiabatic solution. 

When the next COFLASY output (the saved state) becomes  available, the Python driver first attempts continuation using the configured C++ solver profiles in sequence. It starts with the standard settings and proceeds to increasingly permissive fallback settings until the interval is successfully completed. Only if every continuation profile fails does the driver repeat the same profile sequence with fresh-step initialization. In calculations performed at large lepton flavor asymmetries, the numerical complexity of the coupled equations prevents continuation from succeeding for most intervals, so the fresh-step initialization is used instead.

\subsection{Production of sterile neutrinos}
\label{Sec:prod}
We consider a sterile state $\nu_s$ that mixes with an active flavor $\nu_\alpha$ through a mixing angle $\theta_\alpha$. The oscillation probability averaged between collisions is \cite{Venumadhav:2015pla}
\begin{equation}\label{eq:Peff_averaged}
    P_{\mathrm{eff},\alpha}
    =\frac{1}{2}\frac{\Delta_s^2\sin^2 2\theta_\alpha}
    {(\Delta_s\cos 2\theta_\alpha-V_\alpha)^2
    +\Delta^2\sin^2 2\theta_\alpha+(\Gamma_\alpha/2)^2},
\end{equation}
where, for a neutrino with physical 3-momentum $p$, we define
\begin{equation}
    \Delta_s(p)\equiv\frac{m_s^2-m_{\nu_\alpha}^2}{2p}
    \simeq \frac{m_s^2}{2p},
\end{equation}
$\Gamma_\alpha(p,T)$ is the active-neutrino interaction rate, and the matter potential $V_\alpha$ is given by

\begin{equation}
\begin{split}
    \frac{V_\alpha}{\sqrt{2}G_F}
    ={}& \Delta n_{\nu_\alpha}+\Delta n_{l_\alpha} +\sum_{\beta=e,\mu,\tau}\left[
    \Delta n_{\nu_\beta}
    +\left(-\frac{1}{2}+2\sin^2\theta_W\right)
    \Delta n_{l_\beta}\right]\\
    &+(1-2\sin^2\theta_W)\Delta n_\pi  
    -\frac{8p}{3}\left(
    \frac{\rho_{\nu_\alpha}}{m_Z^2}
    +\frac{\rho_{l_\alpha}}{m_W^2}\right).
\end{split}
\end{equation}

In the primordial plasma, a semiclassical kinetic description of sterile-neutrino production is provided by the Boltzmann equation on the sterile neutrino phase space density  
\begin{equation}\label{eq:sterile_boltzmann}
    \left(\frac{\partial}{\partial t}
    -Hp\frac{\partial}{\partial p}\right)f_{\nu_s}(p,t)
    =\frac{\Gamma_\alpha(p,T)}{2}
    P_{\mathrm{eff},\alpha}(p,T)
    \left[f_{\nu_\alpha}(p,T)-f_{\nu_s}(p,t)\right].
\end{equation}

At very large asymmetries, a resonance may be crossed in less than one oscillation period, so a naive oscillation average is not justified. We therefore use the generalized effective-probability prescription of Ref.~\cite{Akita:2025txo}, which was constructed to reproduce the quantum kinetic equations in this regime. For the very small mixing angles relevant here, its resonant probability \eqref{eq:Peff_averaged} reduces to the form 
\begin{equation}\label{eq:Peff_generalized}
    P_{\mathrm{eff},\alpha}
    \simeq \frac{1}{2}\frac{\Delta^2\sin^2 2\theta_\alpha}
    {(\Delta\cos 2\theta_\alpha-V_\alpha)^2
    +(\Gamma_\alpha/2)^2}.
\end{equation}

The Mikheyev--Smirnov--Wolfenstein resonance occurs at each 3-momentum $p\equiv yT$ when the first term in the denominator of eq.\,\eqref{eq:Peff_generalized} turns to zero, 
\begin{equation}\label{eq:sterile_resonance}
    g_\alpha(T,y)\equiv
    \Delta(T,y)\cos 2\theta_\alpha-V_\alpha(T,y)=0,
    \qquad T=T_{\mathrm{res}}(y).
\end{equation}
For antineutrinos, the charge-asymmetric part of $V_\alpha$ changes sign. When the resonance is narrow, the factor in Eq.~\eqref{eq:Peff_generalized} may be safely replaced with $\delta$-function, 
\begin{equation}\label{eq:nwa_delta}
    \frac{\Gamma_\alpha/2}
    {g_\alpha^2+(\Gamma_\alpha/2)^2}
    \longrightarrow \pi\,\delta(g_\alpha).
\end{equation}

Neglecting any backreaction in the r.h.s. of Eq.~\eqref{eq:sterile_boltzmann} associated with presumably small amount of sterile neutrinos, the integration through all the resonance crossings can be performed analytically yielding the following distribution  
\begin{equation}\label{eq:sterile_nwa}
    f_{\nu_s}(y)=\sum_{\alpha}
    \left.
    \left(-\frac{dt}{dT}\right)
    \frac{\pi}{2}\Delta^2\sin^2 2\theta_\alpha\,
    f_{\nu_\alpha}
    \left|\frac{\partial g_\alpha}{\partial T}\right|^{-1}
    \right|_{T=T_{\mathrm{res}}(y)},
\end{equation}
with 
\begin{equation}
    \frac{dt}{dT}=-\frac{d\rho/dT}{3H(\rho+P)},
\end{equation}
where $\rho$ and $P$ are energy density and pressure of the plasma and $H$ is the Hubble parameter. 

For the probability in Eq.~\eqref{eq:Peff_generalized}, the temperature width of the resonance is approximately
\begin{equation}\label{eq:res_width}
    \delta T_{\mathrm{res}}
    \simeq
    \left.
    \frac{\Gamma_\alpha/2}
    {|\partial g_\alpha/\partial T|}
    \right|_{T=T_{\mathrm{res}}}.
\end{equation}
The resonant description requires the crossing to be simple and isolated, $\partial g_\alpha/\partial T\neq0$, and each factor $F(T)$ multiplying the resonant contribution to vary slowly across this interval, that means
\begin{equation}\label{eq:nwa_conditions}
    \delta T_{\mathrm{res}}
    \left|\frac{d\ln F}{dT}\right|_{T_{\mathrm{res}}}\ll1,
    \qquad
    \frac{\delta T_{\mathrm{res}}}{T_{\mathrm{res}}}\ll1.
\end{equation}
Here $F$ refers to $dt/dT$, $\Delta^2$, $f_{\nu_\alpha}$, $\Gamma_\alpha$, and the thermodynamic quantities entering the thermal potentials  $V_\alpha$. The second condition in \eqref{eq:nwa_conditions} equivalently implies  $T_{\mathrm{res}}|\partial g_\alpha/\partial T|\gg\Gamma_\alpha/2$. The integration range must contain the complete resonance, distinct crossings must not overlap, and the non-resonant contribution must remain subdominant. If the analytic solution~\eqref{eq:sterile_nwa} is used without a depletion term, sterile-neutrino production must also have negligible backreaction on the active distribution and flavor asymmetries. The corresponding estimate of  Ref.~\cite{Akita:2025txo} reads
\begin{equation}\label{eq:sterile_backreaction}
    |L_\alpha|\gg
    4.4\times10^{-4}\left(\frac{1\,\mathrm{keV}}{m_s}\right),
\end{equation}
which is satisfied for sterile neutrino masses of keV-scale and interesting values of the flavor asymmetries $|L_\alpha|>0.1$. For a representative point with $|L_e|=0.1$, the full resonant integral and its narrow-width limit agree at the $\mathcal{O}(0.5\%)$ level, while the non-resonant contribution is typically two to three orders of magnitude smaller. Increasing the asymmetry makes the resonance even narrower. The approximation is therefore applicable for the $|L_\alpha|>0.1$ cases studied here.

\section{Results}
\subsection{Flavor-asymmetry evolution}
\label{Sec:asym}
\begin{figure}[htb!]
\centerline{
    \includegraphics[width=0.33\textwidth]{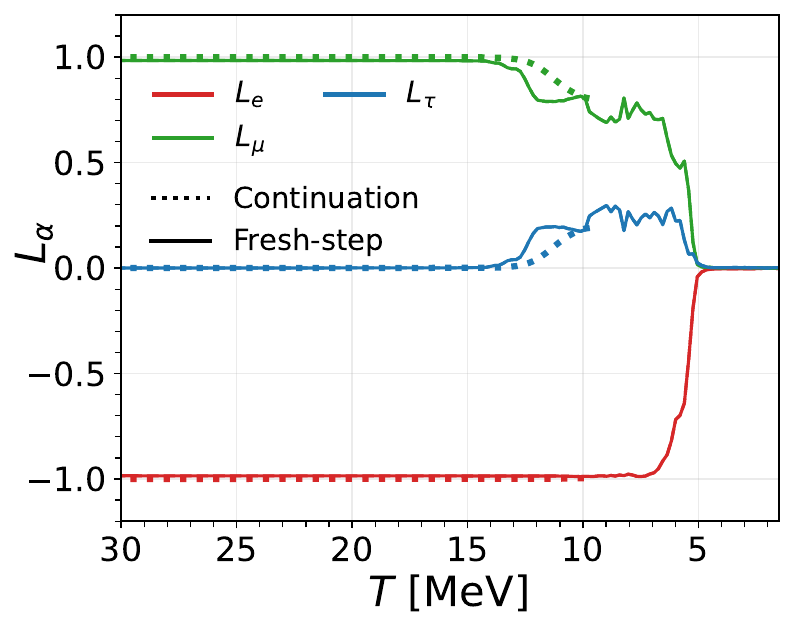}
    \includegraphics[width=0.33\textwidth]{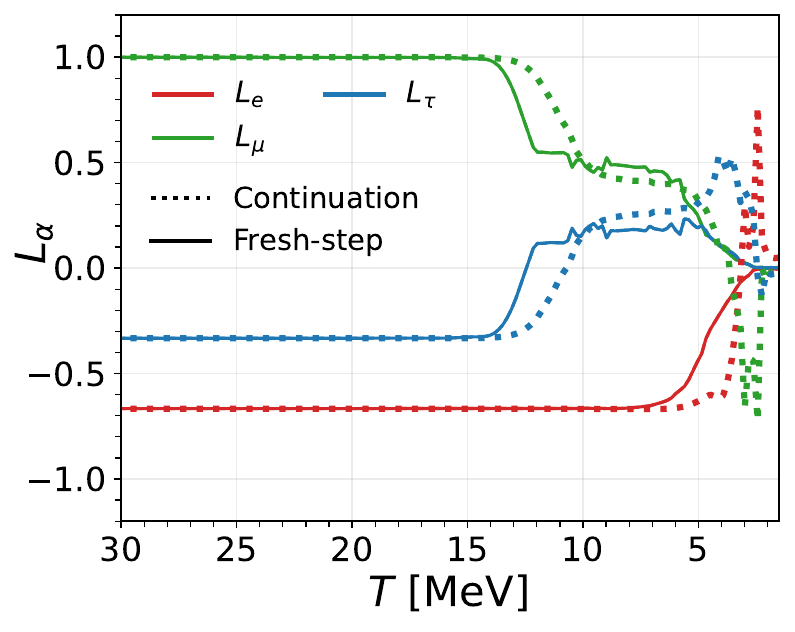}
    \includegraphics[width=0.33\textwidth]{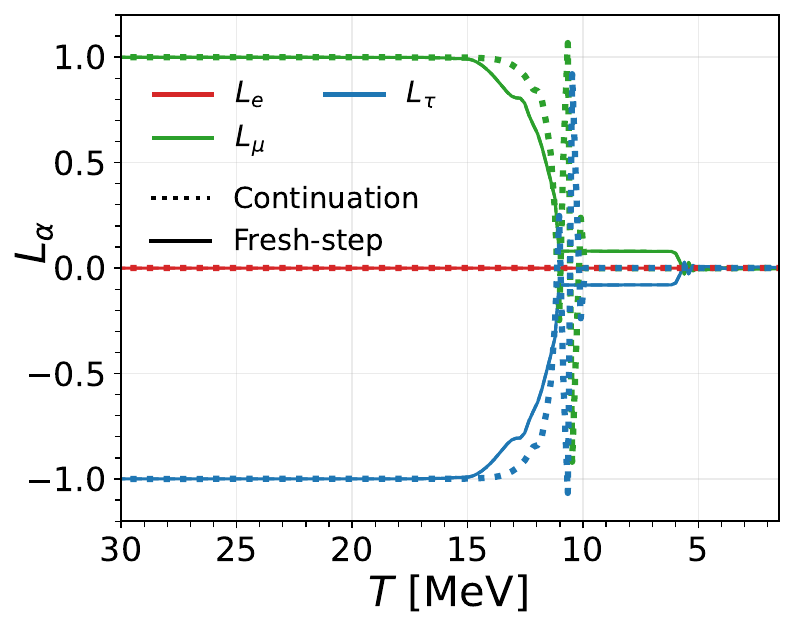}
}
\caption{Lepton-flavor asymmetry evolution with temperature, calculated for three different cases of the initial large lepton flavor asymmetries \eqref{eq:L_vector}: left panel (I), middle panel (II), right panel (III). Solid curves show the fresh-step initialization, while dotted curves show the continuation regime as described in Sec.~\ref{Sec:active}}
\label{fig:BG_cases}
\end{figure}

BBN limits suggest that only the following three classes of large initial flavor asymmetries are allowed \cite{Domcke:2025lzg}:
\begin{equation} \label{eq:L_vector}
\begin{split}
    & [IH]: \quad \quad \; \; L_e = -L_\mu, \; L_\tau = 0 \qquad \qquad \quad (I)\\
    & [NH]: \quad \quad \, L_e = -2/3L_\mu, \; L_\tau = -1/3L_\mu \quad (II) \\
    & [NH,IH]: \; L_\tau = -L_\mu; \; L_e=0 \qquad \quad \quad \; (III)
\end{split}
\end{equation}
where IH and NH denote the inverted and normal neutrino mass hierarchies, respectively. However, a nonzero contribution from charged leptons could modify this conclusion. The three cases in Eq.~\eqref{eq:L_vector} exhibit substantially different background evolutions, as shown in Fig.~\ref{fig:BG_cases}.

Fig.~\ref{fig:BG_cases} also illustrates the different numerical behavior of the two evaluation regimes. Continuation is more computationally demanding because each interval retains the complete dynamical state, including the off-diagonal information accumulated during the preceding evolution. In Case~I, shown in the left panel, the continuation regime fails and cannot be extended to the end of the evolution. In Cases II and III, shown in the middle and right panels, respectively, continuation instead develops anomalous behavior that is absent in the fresh-step solutions. These features indicate a loss of numerical stability in the affected parts of the continuation trajectories. However, such a continuation failure does not necessarily prevent the evaluation of sterile-neutrino production. The sterile spectrum and abundance depend primarily on the background over the temperature interval containing the relevant resonances. If continuation fails at some temperature but the resonance interval is reliably resolved at lower temperatures, the sterile-neutrino spectrum and abundance can still be evaluated from that resolved part of the evolution.

\subsection{Sterile-neutrino abundance}
\label{Sec:abund}
\begin{figure}[htb!]
\centerline{
    \includegraphics[width=0.33\textwidth]{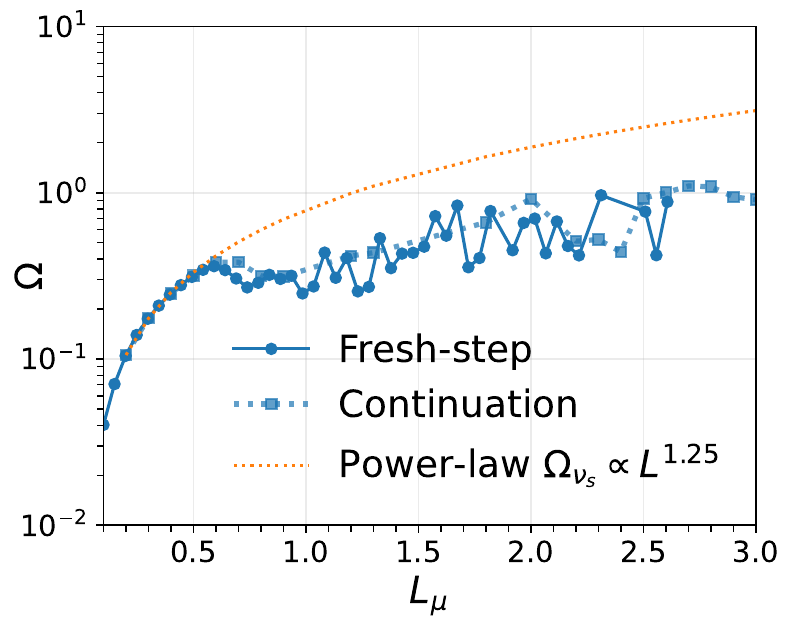}
    \includegraphics[width=0.33\textwidth]{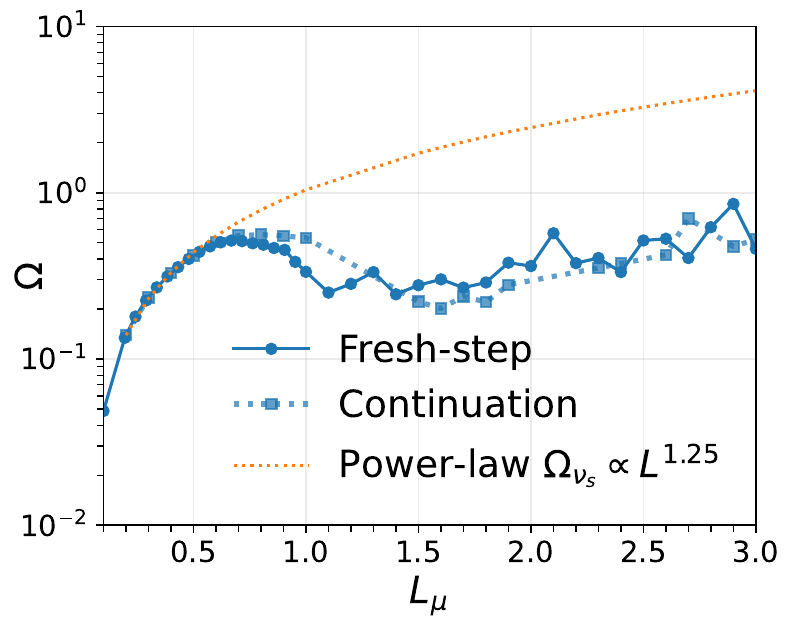}
    \includegraphics[width=0.33\textwidth]{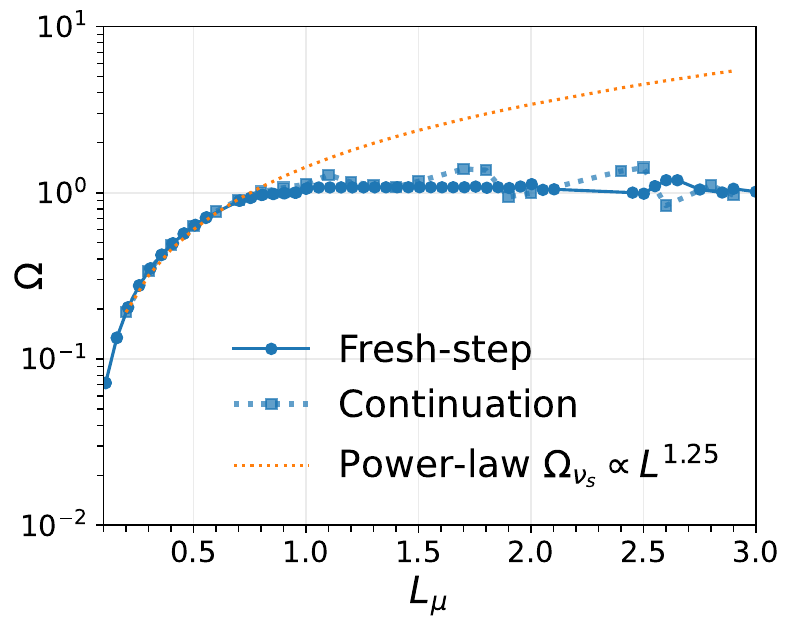}
}
\caption{Sterile neutrino abundances as compared to that of the dark matter, 
$\Omega = \frac{\Omega_{\nu_s}}{\Omega_\text{DM}}$, calculated for sterile neutrino mass $m_s=40$ keV and $\sin^2 2\theta_\mu=2\times10^{-17}$ for three different cases of the initial large lepton flavor asymmetries \eqref{eq:L_vector}: left panel (I), middle panel (II), right panel (III).}
\label{fig:abund_cases}
\end{figure}

For the cases in Eq.~\eqref{eq:L_vector}, Fig.~\ref{fig:abund_cases} shows the resulting abundance as a function of the initial asymmetry for $m_s=40\,\mathrm{keV}$ and the sterile neutrino mixing with muon neutrino governed by the angle $\theta_\mu$ obeying $\sin^2 2\theta_\mu=2\times10^{-17}$. The characteristic features and non-monotonic behavior of these curves arise from the distinct background evolutions associated with the three flavor configurations; their physical origin is discussed in the following subsection.

In Cases I and II, the sudden drop in $\Omega_{\nu_s}$ shown in Fig.~\ref{fig:abund_cases} is primarily associated with an increase in the resonance-slope factor $|\partial g_\alpha/\partial T|$, which enters Eq.~\eqref{eq:sterile_nwa}. Up to the initial asymmetries of approximately $L\simeq0.5$--$0.7$, the resonant production occurs predominantly before the onset of active-flavor oscillations. As the initial asymmetry increases, the resonance interval shifts to lower temperatures and begins to overlap with the rapid decrease of the relevant flavor asymmetry induced by active oscillations. The corresponding rapid evolution of the matter potentials increases $|\partial g_\alpha/\partial T|$ at the resonance, suppressing sterile-neutrino production and producing the pronounced drop in the sterile neutrino abundance. A further increase in the initial asymmetry shifts the resonances into the subsequent asymmetry plateau, where the matter potential evolves slowly. The resonance-slope factor then decreases again, allowing the abundance to resume its growth. At even larger initial flavor asymmetries, the resonance interval reaches the epoch of electron-flavor oscillations near $T\approx 5$ MeV, producing a second drop in the abundance. Unlike the earlier transition these oscillations drive all the flavor asymmetries toward zero. Hence, there is no subsequent plateau with sufficient asymmetry to restore the resonant production, and further increases in the initial asymmetry do not lead to renewed growth of the abundance. At such large initial asymmetries, however, the results are subject to substantial numerical uncertainties.

\subsection{Minimal active--sterile mixing angle}
\label{Sec:minimal-mixing}

The enhancement of resonant production at large lepton asymmetry allows the observed dark-matter density to be obtained with a smaller active--sterile mixing angle. At fixed sterile-neutrino mass and mixing angle, the produced abundance initially increases with the magnitude of the primordial asymmetry. This behavior persists approximately up to $L_\mu\simeq0.5$--$1.0$, with the precise value depending on the flavor configuration and on the type of the active flavor mixed with the sterile state. Equivalently, the minimum value of $\sin^2 2\theta_\alpha$ required to reproduce the observed dark-matter abundance decreases over this range.

\begin{figure}[htb!]
\centerline{
    \includegraphics[width=0.33\textwidth]{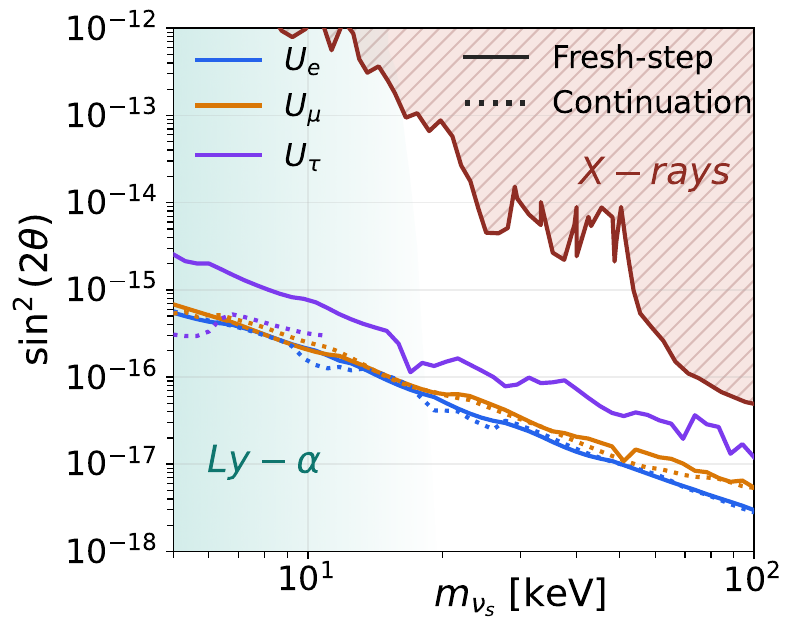}
    \includegraphics[width=0.33\textwidth]{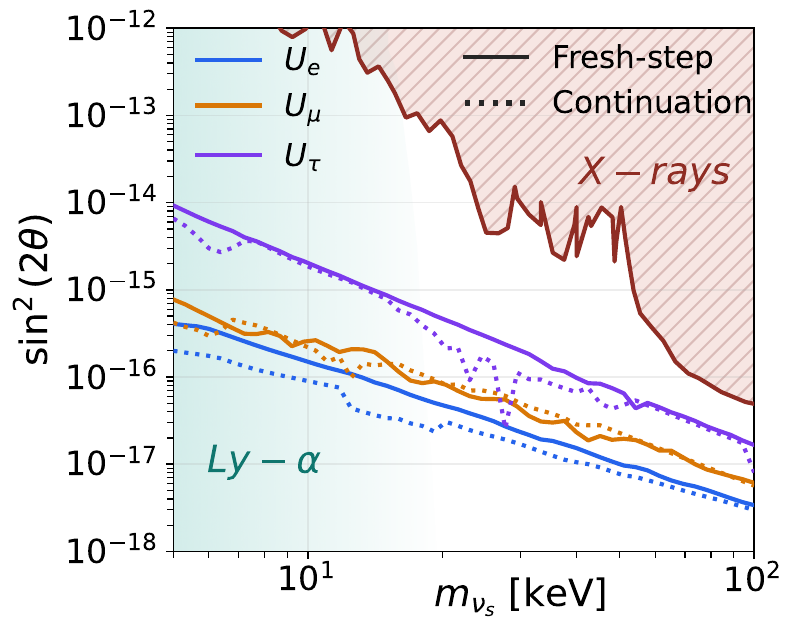}
    \includegraphics[width=0.33\textwidth]{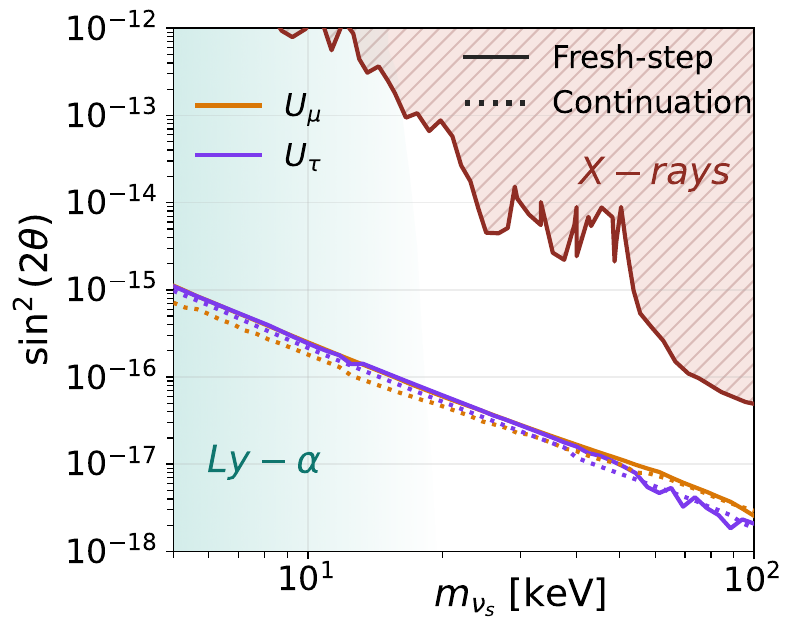}
}
\caption{Minimal active--sterile mixing angles as functions of the sterile-neutrino mass for the three initial flavor-asymmetry configurations\,\eqref{eq:L_vector}: I, II, III correspond to panels from left to right. Solid colored curves show the minimal mixing angles inferred after maximizing the abundance over the initial asymmetry, while dashed curves show the corresponding results at $L_\mu=1$. The shaded regions indicate the  constraints from X-ray telescopes and analyses of Ly-$\alpha$ forests adapted from Ref.\,\cite{Akita:2025txo}.}
\label{fig:minimal_mixing}
\end{figure}

For larger initial asymmetries, the resonance is shifted to lower temperatures, where the active neutrino oscillations efficiently redistribute the flavor asymmetries. The asymmetry entering the relevant active--sterile matter potential is then depleted before or during the resonant production. Consequently, increasing the initial asymmetry no longer produces a comparable increase in the sterile-neutrino density, and the required mixing angle approaches a limiting value. Figure~\ref{fig:minimal_mixing} 
shows the resulting minimal mixing angles for the three flavor-asymmetry configurations in Eq.~\eqref{eq:L_vector}. For each mass and flavor configuration, the minimal mixing angle is determined from the largest value of $\Omega_{\nu_s}$ obtained while varying the initial asymmetry over the range $0.1 \leq L \leq 3$. The value of $L$ at which this maximum occurs depends on both $m_s$ and the flavor configuration in Eq.~\eqref{eq:L_vector}, owing to the dependence of the resonance temperature $T_{res}$ on the sterile-neutrino mass $m_s$.

\subsection{Sterile-neutrino momentum spectra}
\label{sec:sterile_spectra}

\begin{figure}[htb!]
\centerline{
    \includegraphics[width=0.33\textwidth]{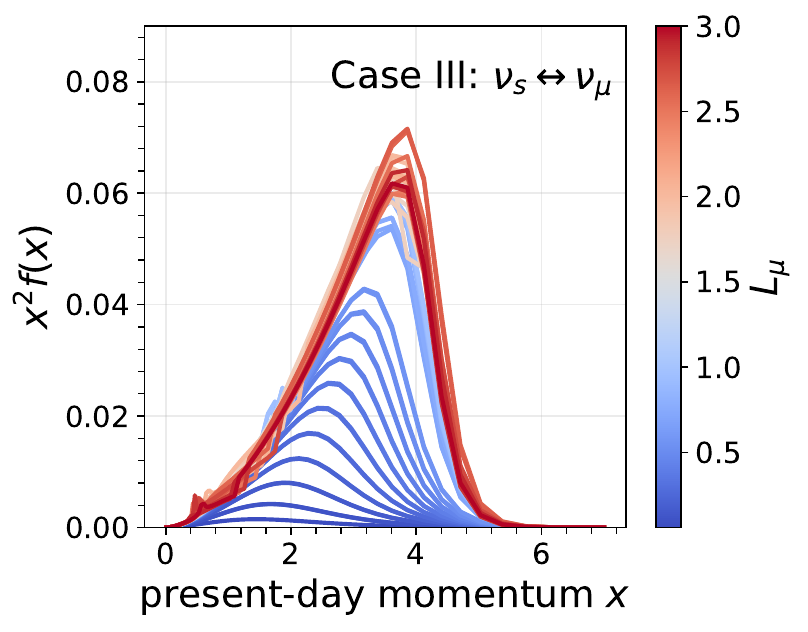}
    \includegraphics[width=0.33\textwidth]{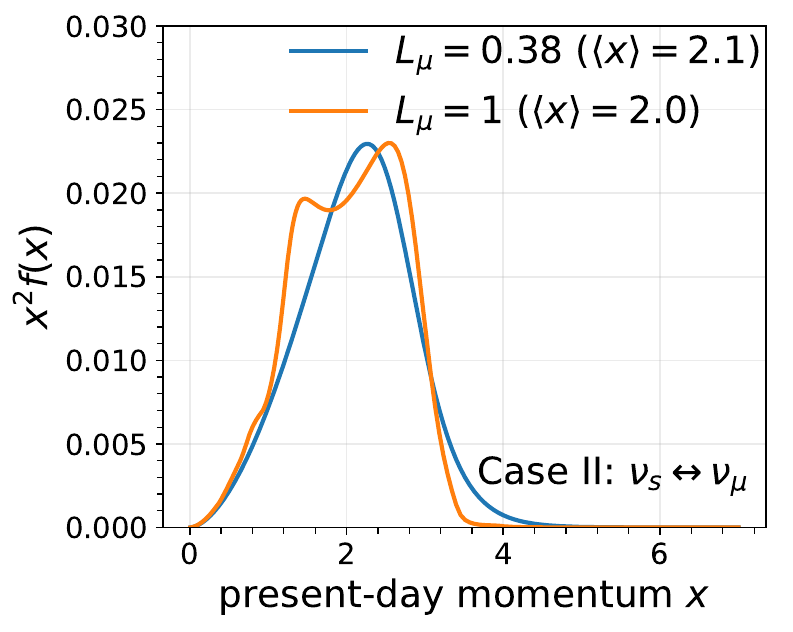}
    \includegraphics[width=0.33\textwidth]{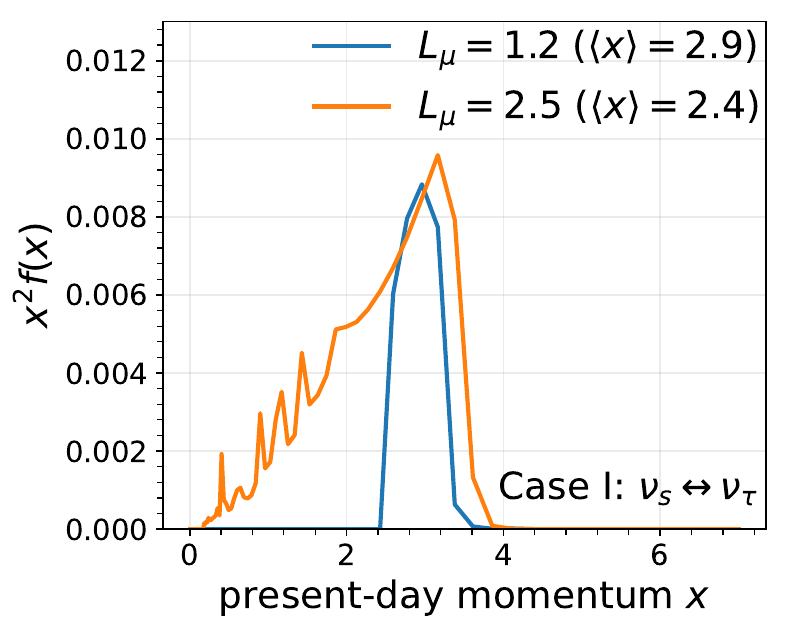}
}
\caption{Present-day sterile-neutrino number spectra for $m_s=40\,\mathrm{keV}$ in the fresh-step regime, sterile-neutrino momenta are normalized to the present day temperature of relic photons, $p=x T_\gamma$.}
\label{fig:sterile_spectra}
\end{figure}

The momentum distributions provide further information about resonant production and its implications for structure formation. Left panel of the Fig.~\ref{fig:sterile_spectra} shows the present-day number spectra for Case~III with muon-flavor mixing and $m_s=40\,\mathrm{keV}$. As the initial asymmetry grows, the distribution shifts toward higher momenta. Hereafter we characterize it via ratio to the present temperature of cosmic microwave background, $x\equiv p/T_\gamma$, $T_\gamma=2.725$\,K. With increasing asymmetry the average momentum also increases with $\langle x\rangle$ approaching approximately $3.2$. The resulting hotter dark matter is constrained by structure formation. The adopted Ly-$\alpha$ bound \cite{Akita:2025txo}, shown in Fig.~\ref{fig:minimal_mixing}, is based on a simple one-parametric (average momentum) fit and excludes sterile-neutrino masses below approximately $20\,\mathrm{keV}$.

However, the mean momentum alone can not characterize the impact of a nonthermal dark matter distribution on structure formation. In particular, the unusual spectral shapes that arise when sterile production overlaps with active-flavor oscillations require an analysis using the full momentum distribution.

\begin{figure}[htb!]
  \centering
    \includegraphics[width=0.8\textwidth]{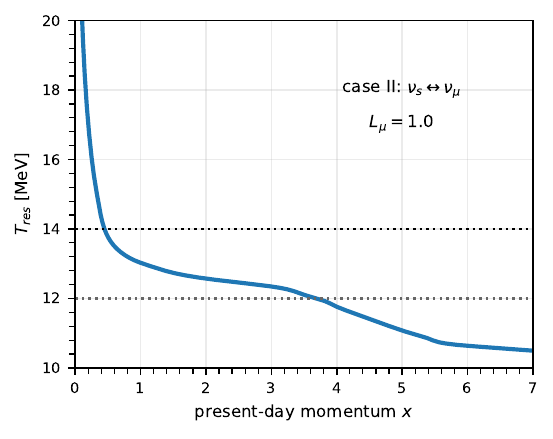}
  \caption{Resonant temperatures for the present day momentum $x$ at $L_\mu=1.0$ in Case II for $\nu_s \leftrightarrow \nu_\mu$ mixing and $m_{\nu_s}=40$ keV.}
  \label{fig:resonance_vs_momentum}
\end{figure}

An example is shown in the middle panel of Fig.~\ref{fig:sterile_spectra} for Case~II with muon-flavor mixing. The two initial asymmetries, $L_\mu=0.38$ and $L_\mu=1$, yield similar relic abundances and mean momenta, but considerably different spectral shapes. The twin peaks distribution at $L_\mu=1$ reflects the complicated regime of production. Figure \ref{fig:resonance_vs_momentum} 
illustrates how different momentum modes probe successive stages of this evolution. The lowest-momentum modes encounter the resonance earlier, at higher temperatures, while the relevant flavor asymmetry is still large. During active oscillations, the rapid variation of the matter potential increases the resonance-slope factor $|\partial g_\alpha/\partial T|$, whose inverse enters Eq.~\eqref{eq:sterile_nwa}, thereby suppressing production during the transition. For $x \gtrsim 2$, the resonance temperature varies only slowly with momentum: modes spanning approximately $2 \lesssim x \lesssim 3$ are produced within the narrow interval. At still larger momenta, the resonances probe the subsequent asymmetry plateau, shown in the middle panel of Fig.~\ref{fig:BG_cases}. The reduced asymmetry then makes resonant production inefficient, suppressing the high-momentum tail.

Particularly interesting and computationally demanding spectra arise in Case~I with tau-flavor mixing, illustrated on the right panel of Fig.~\ref{fig:sterile_spectra}. Although the initial tau-flavor asymmetry is zero, active oscillations generate a nonzero $L_\tau$ for a finite interval of $T$. Resonant sterile production can then occur if the resonance temperatures fall within this interval and the induced asymmetry is sufficiently large. For the smaller initial asymmetry shown, $L_\mu=1.2$, these conditions are satisfied only over a narrow momentum range, yielding a sharply localized spectrum. At the larger initial asymmetry, $L_\mu=2.5$, production extends to lower momenta. This low-momentum component is nevertheless suppressed because it is produced while the tau-flavor asymmetry is still building up.

For these special cases at large asymmetry, particularly Case~I with tau-flavor mixing, the present numerical uncertainties prevent a reliable determination of the detailed momentum distribution. The examples illustrate the qualitative production mechanism, but small-scale irregularities in the calculated spectra should not be assigned a physical interpretation without further convergence checks. Deriving robust structure-formation constraints for these cases requires improvements in the numerical treatment and mathematical control of the overlapping active-oscillation and sterile-production epochs, followed by an analysis of structure formation using the resolved spectra.

\section{Conclusion}

In this work, we investigated the resonant sterile-neutrino dark-matter production while accounting for the evolution of large lepton-flavor asymmetries. We find that increasing the initial asymmetry does not necessarily lead to an unlimited enhancement of the sterile-neutrino abundance. A larger asymmetry shifts resonant production to lower temperatures, where the active-neutrino oscillations become efficient and redistribute or deplete the flavor asymmetry that generates the resonance. The resulting interplay provides a saturation mechanism: beyond a flavor- and mass-dependent range of initial asymmetries, a further increase in the primordial asymmetry produces little or no increase in the final sterile-neutrino abundance. Whether the required large asymmetries may be generated within particular realistic models of sterile neutrino dark matter, like e.g. $\nu$MSM \cite{Asaka:2005an,Asaka:2005pn,Canetti:2012kh}, remains an open question to be further investigated. 

A precise determination of the maximal abundance and the associated momentum spectra remains limited by the numerical complexity of the coupled evolution, particularly in the regime where active-flavor conversion and sterile-neutrino production occur over overlapping temperature intervals. Improving this calculation is therefore an important direction for future work. In particular, the background evolution and the active-neutrino quantum kinetic equations should be integrated within a single continuous numerical framework, rather than coupled through separate temperature steps. Such a unified treatment, together with improved numerical stability and resolution, will be required to determine the maximal sterile-neutrino abundance and momentum distribution with greater precision. It is also required for the accurate calculation of the spectra of produced sterile neutrino dark matter. Although the presently allowed model parameter space favors heavier sterile neutrinos, with larger asymmetries and later production the produced particles are generically hotter\,\cite{Gorbunov:2025nqs} and hence specifically affect the formation of cosmic large scale structure. The absence of the corresponding features in galaxy counting \cite{Newton:2018izu,Newton:2020cog}, Ly-$\alpha$ forest \cite{Villasenor:2022aiy,Vogel:2025aut}, phase space density of galactic dark matter \cite{Bezrukov:2025ttd,Alvey:2020xsk} etc, allow one to exclude models with lighter sterile neutrinos. 

Our calculations show that at least an order of magnitude improvement in the presently achieved sensitivity to the active-to-sterile mixing angle is needed to fully explore the resonance production mechanism with future $X$-ray telescopes aiming at searches for the peak-like feature in the spectra of galaxies and galaxy clusters. 

An accurate treatment of the pion condensate is also necessary for a precise calculation of sterile-neutrino production. Its contribution to charge neutrality and the thermodynamic properties of the plasma modifies the chemical potentials and background evolution that determine the resonant conversion. More generally, large primordial lepton-flavor asymmetries can alter the cosmological trajectory through the QCD phase diagram. In the quark--meson model of Ref.~\cite{Ferreira:2025zeu}, the entry into the pion-condensed phase can occur through a first-order phase transition, accompanied with production of relic gravitational waves. A future detection attributable to this QCD-era mechanism could provide independent evidence for a large primordial lepton asymmetry and thereby support the resonant sterile-neutrino dark-matter scenario studied here.

\section*{Acknowledgments}

This work is supported in the
framework of the State project “Science” by the Ministry of Science and Higher Education
of the Russian Federation under the contract 075-15-2024-541.

\bibliographystyle{unsrt}
\bibliography{refs}

\end{document}